\documentclass[doublecol,final]{epl2} 

\usepackage{amssymb}
\usepackage{amsmath}
\usepackage{amsfonts}
\usepackage{graphics}
\usepackage{float}

\renewcommand{\v}[1]{\mathbf{#1}}
\newcommand{\twovec}[2]{\ensuremath{\begin{pmatrix}#1\\#2\end{pmatrix}}}

\title{Relaxation Dynamics of Half-Quantum Vortices in a Two-Dimensional Two-Component Bose-Einstein Condensate}
\shorttitle{Relaxation Dynamics of Two-Component BEC} 

\author{M.~T.\ Wheeler$^{\dag}$ \and H.\ Salman$^{\star}$ \and M.~O.\ Borgh$^{\dag}$}
\shortauthor{M.~T.\ Wheeler \etal}

\institute{
$^{\dag}$Physics, Faculty of Science, University of East Anglia, NR4 7TJ, United Kingdom\\
$^{\star}$School of Mathematics, University of East Anglia, NR4 7TJ, United Kingdom\\
}

\abstract{We study the relaxation dynamics of quantum turbulence in a two-component Bose-Einstein condensate containing half-quantum vortices. We find a temporal scaling regime for the number of vortices and the correlation lengths that at early times is strongly dependent on the relative strength of the inter-species interaction. At later times we find that the scaling becomes universal, independent of the inter-species interaction, and approaches that numerically observed in a scalar Bose-Einstein condensate.}

\begin{document}

\maketitle

\section{Introduction}
Since the realization of superfluidity, quantum turbulence (QT) has been studied in systems ranging from superfluid liquid Helium~\cite{barenghi-PNAS-2014,walmsley-PNAS-2014} to quasi-particle condensates in solid-state systems~\cite{kreil-PRL-2018}. Due to their unprecedented experimental accessibility, QT in Bose-Einstein condensates (BECs) in dilute, ultracold atomic gases have attracted considerable theoretical~\cite{kobayashi-PRA-2007,numasato-PRA-2010, reeves-PRL-2013,billam-PRL-2014,simula-PRL-2014,baggaley-PRA-2018} and experimental~\cite{henn-PRL-2009,kwon-PRA-2014,seo-SciRep-2017,navon-Science-2019,guillaume-Science-2019,johnstone-Science-2019} interest in both 2D and 3D configurations. In a scalar BEC, the QT state is made up of a large number of vortices with quantised circulation. The collective behaviour of the vortices plays a key role in the hydrodynamics, recovering features of classical turbulence that can exhibit the characteristic Kolmogorov power-law spectrum~\cite{kobayashi-JPS-2005}. \par

In contrast to the scalar superfluids, multicomponent and spinor BECs are described by multicomponent order parameters and allow for a wider range of topological defects, which give rise to novel dynamics~\cite{kasamatsu-PRA-2016,weiss-NComms-2019,kobayashi-PRL-2009,kasamatsu-IJP-2005}.
Consequently, there has been increasing interest in the properties of QT and non-equilibrium dynamics in such systems~\cite{salman-PhysicaD-2009,schmied-PRA-2019, karl-PRA-2013, prufer-Nature-2018,hofmann-PRL-2014}.  
The simplest non-scalar topological excitation appears in a two-component BEC, described by two complex fields, as the appearance of a phase singularity in only one component. 
When the atomic mass and mean density of the components are equal, such vortices are often referred to as half-quantum vortices (HQVs), due to their similarities with vortices carrying half a quantum of superfluid circulation in superfluid $^3$He~\cite{volovik-JETPL-1976,autti-PRL-2016} and spin-1 BECs~\cite{leonhardt-JETPL-2000,seo-PRL-2015}.
The study of QT in BECs can be separated into two distinct categories: 1) forced turbulence where a statistically stationary state is established; 2) decaying turbulence where a non-equilibrium initial condition, typically involving vortices, relaxes towards equilibrium. 
Here, we numerically investigate the spatial and temporal properties of the relaxation dynamics of a non-equilibrium initial state in a two-dimensional two-component system containing HQVs. 
Using a pseudospin interpretation, we compute the temporal scaling of the correlation functions associated with the spin- and mass-superfluid ordering. 
We relate these to the vortex decay rate and analyse how this depends on the intra-component interaction strength.
We contrast our observations for this system with similar simulations that have been performed for scalar BECs and reported in ~\cite{schole-PRA-2012,nowak-PRA-2012,karl-NJP-2017}.

\section{The two-component BEC}
We consider an untrapped two-component BEC described by the Gross-Pitaevskii (GP) mean-field theory subject to periodic boundary conditions. The dynamics of the condensate is described by the two coupled GP equations

\begin{align}
\label{eq:dimful-GPEs}
\begin{aligned}
    i\hbar \frac{\partial \psi_{1,2}}{\partial t} &= \left(-\frac{\hbar^2}{2m_{1, 2}}\nabla^2 + g_{1,2}|\psi_{1,2}|^2 + g_{12}|\psi_{2,1}|^2\right)\psi_{1,2}
\end{aligned}
\end{align}

\hspace{-0.2cm} where $\psi_j$ is the condensate wavefunction and $m_j$ $(j=1,2)$ is the atomic mass for the $j$th component. The strength of inter- and intra-component interactions are described by $g_j$ and $g_{12}$, respectively. We consider a condensate where $m_1=m_2=m$, as is the case, e.g., when the two components are different hyperfine states of the same atomic species, and also assume $g_1=g_2=g$. The key parameter is then the ratio of intra- to inter-species interactions
\begin{align}
    \gamma = \frac{g_{12}}{g}, 
\end{align}
which in experiment could be tuned using magnetic~\cite{inouye-nature-1998} or microwave-induced~\cite{papoular-PRA-2010} Feshbach resonances. Here we consider $0 < \gamma < 1$, such that all interactions are repulsive, while keeping the condensate stable against separation of the components.

The vortex states of the two-component BEC may be understood as follows: We write the two-component wavefunction as the vector $(\psi_1,\psi_2)^T$. Taking $\theta_j=\mathrm{Arg}(\psi_j)$, this may be decomposed as
\begin{align}
  \label{eq:pseudospinor}
  \twovec{\psi_1}{\psi_2} = \twovec{|\psi_1|e^{i\theta_1}}{|\psi_2|e^{i\theta_2}} = e^{i\Theta}\twovec{|\psi_1|e^{i\Phi}}{|\psi_2|e^{-i\Phi}},
\end{align}
where
\begin{align}
    \Theta = (\theta_1 + \theta_2) / 2, \qquad \Phi = (\theta_1-\theta_2)/2.
\end{align}
Gradients in $\Phi$ can then be interpreted in terms of pseudospin currents, while gradients in $\Theta$ may be associated with a total, superfluid mass current.

Now consider a vortex state consisting of a phase singularity in $\psi_1$, around which $\theta_1$ winds by $2\pi$ while $\theta_2$ remains unchanged, such that
\begin{align}
  \twovec{\psi_1}{\psi_2} = \twovec{|\psi_1|e^{i\phi}}{|\psi_2|} = e^{i\phi/2}\twovec{|\psi_1|e^{i\phi/2}}{|\psi_2|e^{-i\phi/2}},
\end{align}
where $\phi$ is the azimuthal angle around the vortex. The vortex is thus equivalently described by a $\pi$ change in $\Theta$ (and a simultaneous $\pi$ change in $\Phi$) along a closed path encircling the vortex. Since $\Theta$ can be associated with a total mass current in the two components together, these vortex states are often referred to as HQVs and we adopt this language from here on. However, the two-component vortices are topologically distinct from HQVs in the $A$ and polar phases of superfluid $^3$He~\cite{volovik-JETPL-1976,autti-PRL-2016} and in the uniaxial nematic phase of spin-1 BECs~\cite{leonhardt-JETPL-2000,seo-PRL-2015}.

A pseudospin picture also allows us to understand the size of HQV cores in terms of an energetic hierarchy of length scales arising from the inter- and intra-component interactions. These length scales are associated, respectively, with variations of the total superfluid density and of the density difference between the components. We thus define the density and spin healing lengths as~\cite{eto-PRA-2011}
\begin{align}
    \label{eq:healing_lengths}
    \xi_d = \frac{\hbar}{\sqrt{2mgn_0}}, \qquad \xi_s = \xi_d \left( \frac{1 + \gamma}{1 - \gamma} \right)^{1/2} \, ,
\end{align}
where $n_0$ is the number density of each component in a uniform system. Since a HQV consists of a phase singularity in only one condensate component, the remaining component is free to fill the vortex core. This can be interpreted as a variation of the pseuodspin $z$-component, whose size is determined by the spin healing length. When $\xi_s\gtrsim\xi_d$, the vortex core can thus expand, lowering the total energy. Therefore, $\gamma$ directly determines the sizes of the vortex cores in the system. A similar energetic hierarchy of length scales leads to dramatic defect-core deformations in spinor BECs~\cite{ruostekoski-PRL-2003}, including splitting of singly quantised vortices into HQVs~\cite{lovegrove-PRA-2012,seo-PRL-2015}.

\section{Numerical method}
To study the dynamics of vortices in a turbulent regime we numerically evolve the time-dependent two-component Gross-Pitaevskii equations using a split-step algorithm~\cite{javanainen-JPA-2006}. We write eq.~(\ref{eq:dimful-GPEs}) in terms of the dimensionless variables: $\tilde{\vect{r}} = \vect{r}/a_s$, $\tilde{t} = t/\tau$, $\tilde{g} = 2mg/\hbar^2$ and $\tilde{\psi_j}=a_s\psi_j$, where $a_s$ is the lattice spacing and $\tau = 2ma_s^2/\hbar$ is the lattice time. The resulting equations then become
\begin{align}
\label{eq:dimless_GPEs}
        i \frac{\partial \psi_{1,2}}{\partial t} = \left(-\nabla^2 + g|\psi_{1,2}|^2 + \gamma g|\psi_{2,1}|^2\right)\psi_{1,2},
\end{align}
where we have dropped the tildes for notational convenience.
Our simulations were performed on a periodic domain of non-dimensional area $L^2$ with side length $L=N_s$ where $N_s^2$ is the number of grid points. We solve eq.~(\ref{eq:dimless_GPEs}) on a grid of $1024^2$ points with $a_s=1$. Motivated by similar work in a scalar BEC~\cite{karl-NJP-2017}, we take $N=3.2\times10^9$ atoms per component and dimensionless $g=L^2/4N$. The non-dimensional density healing length is thus fixed at $N_s/(gN)^{1/2}=2$. 
We now explore the role of the inter-component interaction by varying $\gamma$ within the range $0 < \gamma < 1$.  

\begin{figure}
    \centering
    \includegraphics[width=\columnwidth]{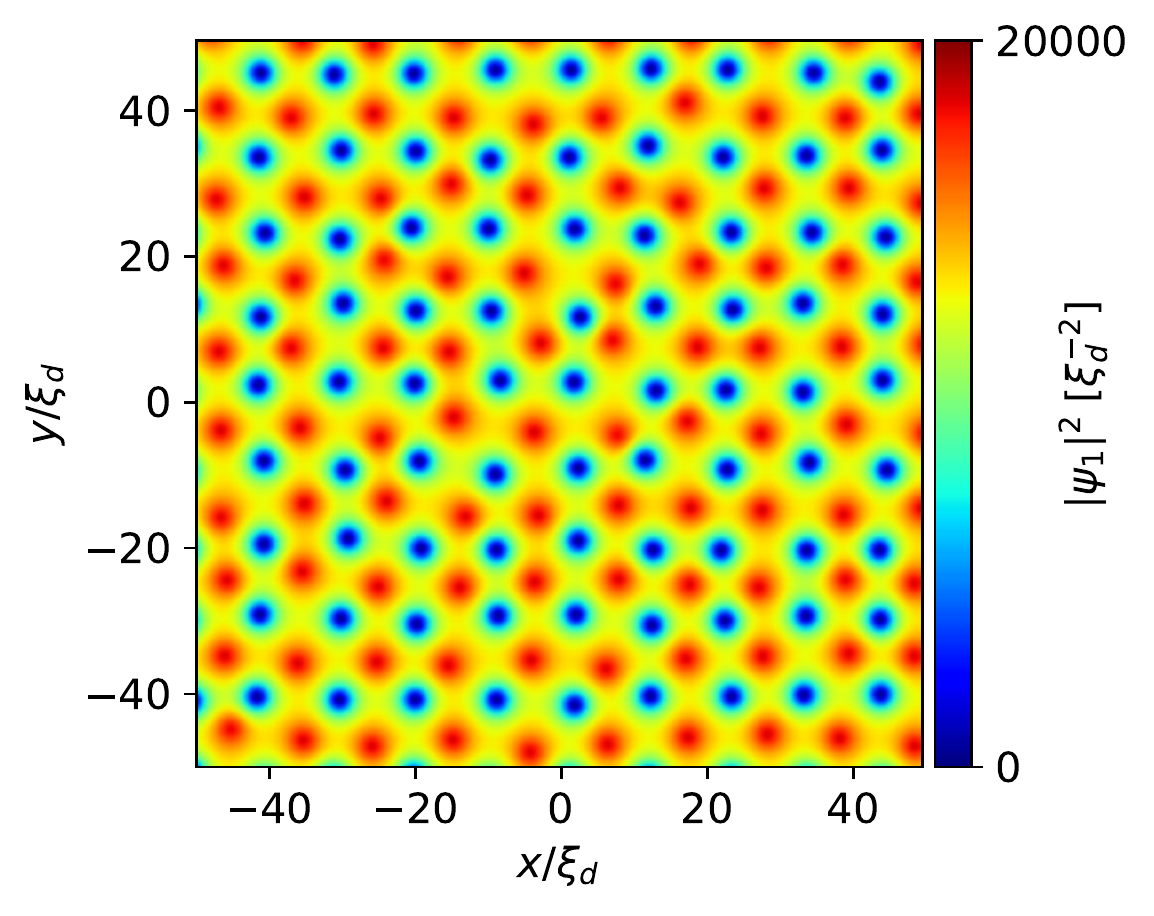}
    \caption{Density $|\psi_1|^2$ in a $100\xi_d \times 100\xi_d$ subdomain of the initial state after a short imaginary-time evolution. We can identify the vortices in this component by the density depletion (blue). Density peaks (red) form in $\psi_1$ at the positions of vortices in the $\psi_2$ component.}
    \label{fig:init_state}
\end{figure}

The initial condition for the GP evolution is constructed as a grid of vortex positions containing $48^2$ vortices in each component with the grids of each component offset in both $x$ and $y$ to avoid overlapping positions. We then add a small, random displacement to each position to create an irregular distribution of vortices. This facilitates the development of an initially chaotic and subsequently turbulent vortex evolution during the relaxation dynamics.
The phase of each component is subsequently constructed as an alternating $2\pi$ winding around each vortex position using the method described in ref.~\cite{reeves-PRL-2013} that also accounts for the periodic boundary conditions. An initial short period of imaginary-time propagation, keeping the phase profile fixed, allows the vortex cores to form. The resulting HQVs consist of a density depletion in one component at the position of the phase singularity, which is then filled with atoms of the other component, as illustrated in fig.~\ref{fig:init_state}. From this initial state, the system is evolved according to eq.~(\ref{eq:dimless_GPEs}). HQVs with opposite circulation but with the phase singularity in the same component may annihilate which leads to a decay of the total vortex number.

\section{Results}
\begin{figure*}[t!]
    \centering
    \includegraphics[width=2\columnwidth]{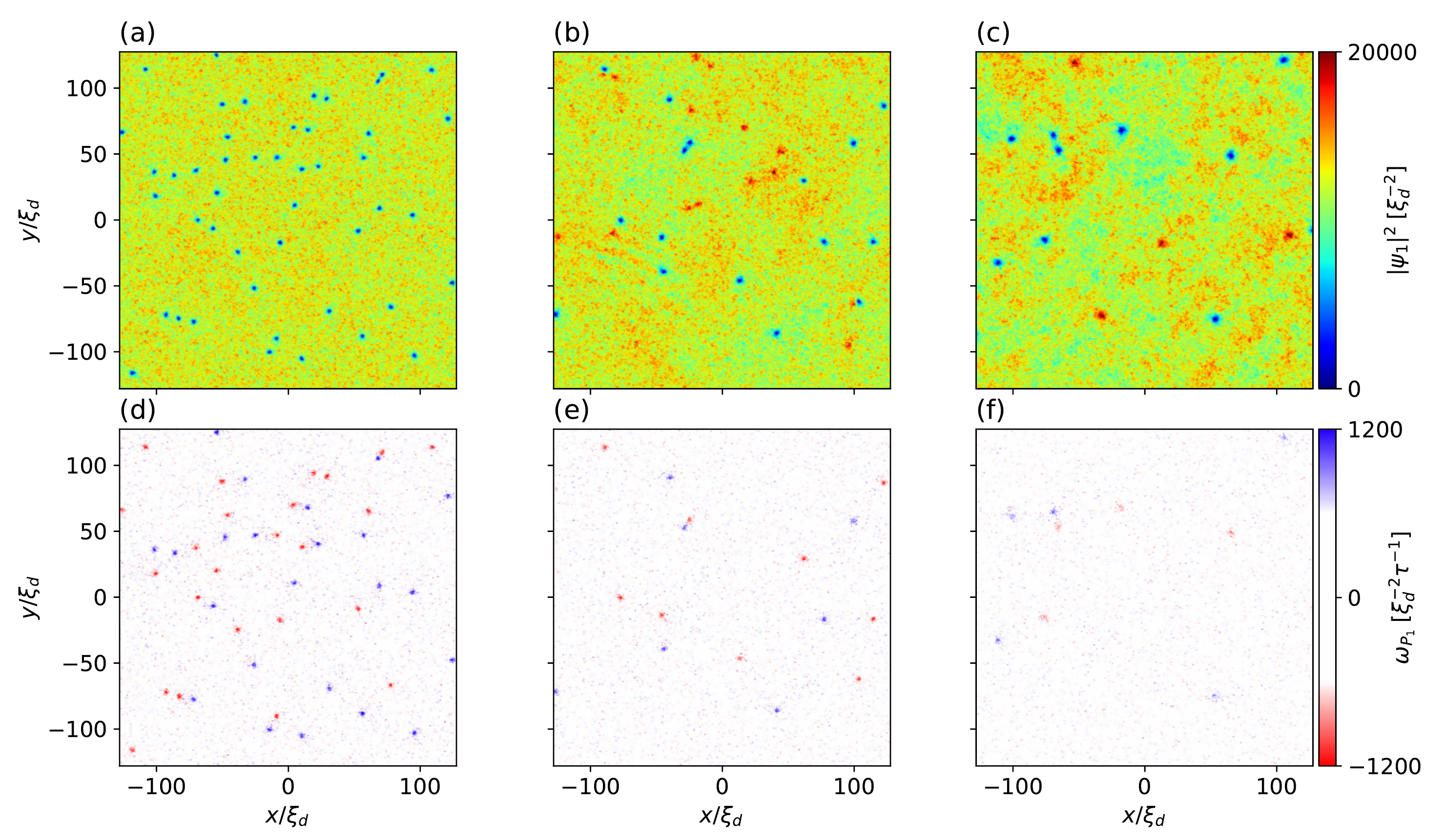}
    \caption{Density (a)--(c) and pseudo-vorticity (d)--(f) of the $\psi_1$ component in a $256\xi_d \times 256\xi_d$ subregion at time $t=2.5\times 10^4\xi_d^2$, for $\gamma=0.1$ (left), $\gamma=0.6$ (middle) and $\gamma=0.8$ (right). Vortices in $\psi_1$ appear as a density depletion. For $\gamma \geq 0.6$, bright density peaks show where $\psi_1$ atoms fill the cores of HQVs with the phase singularity in $\psi_2$. Vortices with positive (blue) and negative (red) circulation are identifiable in the pseudo-vorticity field.}
    \label{fig:densVort}
\end{figure*}
We first investigate the effect of $\gamma$ on the relaxation dynamics of HQVs. 
Fig.~\ref{fig:densVort}(a)--(c) shows the density of the $\psi_1$ component for $\gamma=0.1, 0.6, 0.8$. HQVs with a phase singularity in this component are readily apparent by the corresponding density depletion, and have a core size that grows with increasing $\gamma$. For $\gamma \geq 0.6$, high density peaks also become noticeable and correspond to the positions of HQVs with phase singularity in $\psi_2$. This can be understood from the healing lengths, eq.~\eqref{eq:healing_lengths}.
For small $\gamma$, $\xi_s\sim\xi_d$. As $\gamma$ increases, the spin healing length also increases. Consequently, the cores of the HQVs fill with atoms from the other component as the resulting lowering of the kinetic energy offsets the cost in interaction energy. This causes the vortex cores to expand to a size similar to the spin healing length, as borne out by our simulations.

To track the vortex positions, we evaluate the pseudo-vorticity~\cite{villois-JOP-2016,rorai-JFM-2016}
\begin{align}
    \vect{\omega}_\mathrm{p_j} = \frac{1}{2}\nabla \times (n\vect{v})_j \, ,
\end{align}
where
\begin{align}
  \label{eq:mass_current}
  (n\vect{v})_j = \frac{1}{i} \left[ \psi_j^*(\nabla\psi_j)
  - (\nabla\psi_j^*)\psi_j\right],
\end{align}
is the mass current of component $j=1,2$. The pseudo-vorticity remains regular and non-zero within the core of each vortex, and relaxes to zero away from the vortex singularity (at length scales exceeding the spin-healing length, $\xi_s$), as shown in fig.~\ref{fig:densVort}(d)--(f). The sign of the pseudo-vorticity also determines the charge of the vortex. The pseudo-vorticity shows the vortex positions particularly sharply for small $\gamma$, where the vortex cores are small.


We now investigate the spatial properties of our turbulent system.
We split the kinetic energy, $E_\mathrm{kin} = E^v + E^q$, into  classical ($E^v$), and quantum-pressure ($E^q$) contributions. These are given by
\begin{align}
  \label{eq:E_v}
  E^v &= \frac{1}{4}\int \upd^2\v{x} \, \left( |\sqrt{n}_1 \vect{v}_1|^2 + |\sqrt{n}_2 \vect{v}_2|^2 \right) \, , \\
  \label{eq:E_q}
  E^q &= \int \upd^2\v{x} \, \left( |\nabla \sqrt{n}_1|^2 + |\nabla \sqrt{n}_2|^2 \right) \, , 
\end{align}
where $n_j = |\psi_j|^2$ for $j=1,2$.


The energy spectra for these contributions involve the Fourier transforms
of the generalised velocities for the incompressible ($i$), compressible ($c$), and quantum pressure ($q$) parts~\cite{schmied-PRA-2019}, defined as
\begin{align}
\begin{gathered}
  \label{eq:gen_vel}
    \vect{w}^{i, c} = \sqrt{n_1}\vect{v}_1^{i, c}+\sqrt{n_2}\vect{v}_2^{i, c} \, ,  \\ 
    \vect{w}^q = 2\left(\nabla \sqrt{n_1}+\nabla\sqrt{n_2} \right)\, . 
\end{gathered}
\end{align}
The incompressible and compressible components of the velocity field are recovered from a Helmholtz decomposition into a divergence free, incompressible part $\nabla \cdot \vect{v}^i=0$, and an irrotational, compressible part $\nabla \times \vect{v}^c=\vect{0}$. The kinetic energy spectrum can then be calculated by integrating the corresponding Fourier transforms over the full $k$-space angle 
\begin{align}
  E^{\delta}(k) &= \frac{1}{4} \int_0^{2\pi} \upd \Omega_{\vect{k}} \,
  |\tilde{\vect{w}}^\delta(\vect{k})|^2, \enskip (\delta = i, c, q), 
\end{align}
for wave number $k=|\vect{k}|$.
The total kinetic energy is given by integrating over all $k$
and summing over the different contributions: $E_\textrm{kin} = \sum_{\delta} \int \upd k E^{\delta}(k)$ for $\delta = (i, c, q)$. The occupation numbers corresponding to
the different energy contributions are then
\begin{align}
  \label{eq:occupation_nums}
  n^{\delta}(k) = k^{-2}E^{\delta}(k), \enskip (\delta = i, c, q).
\end{align}

\begin{figure}
    \centering
    \includegraphics[width=\columnwidth]{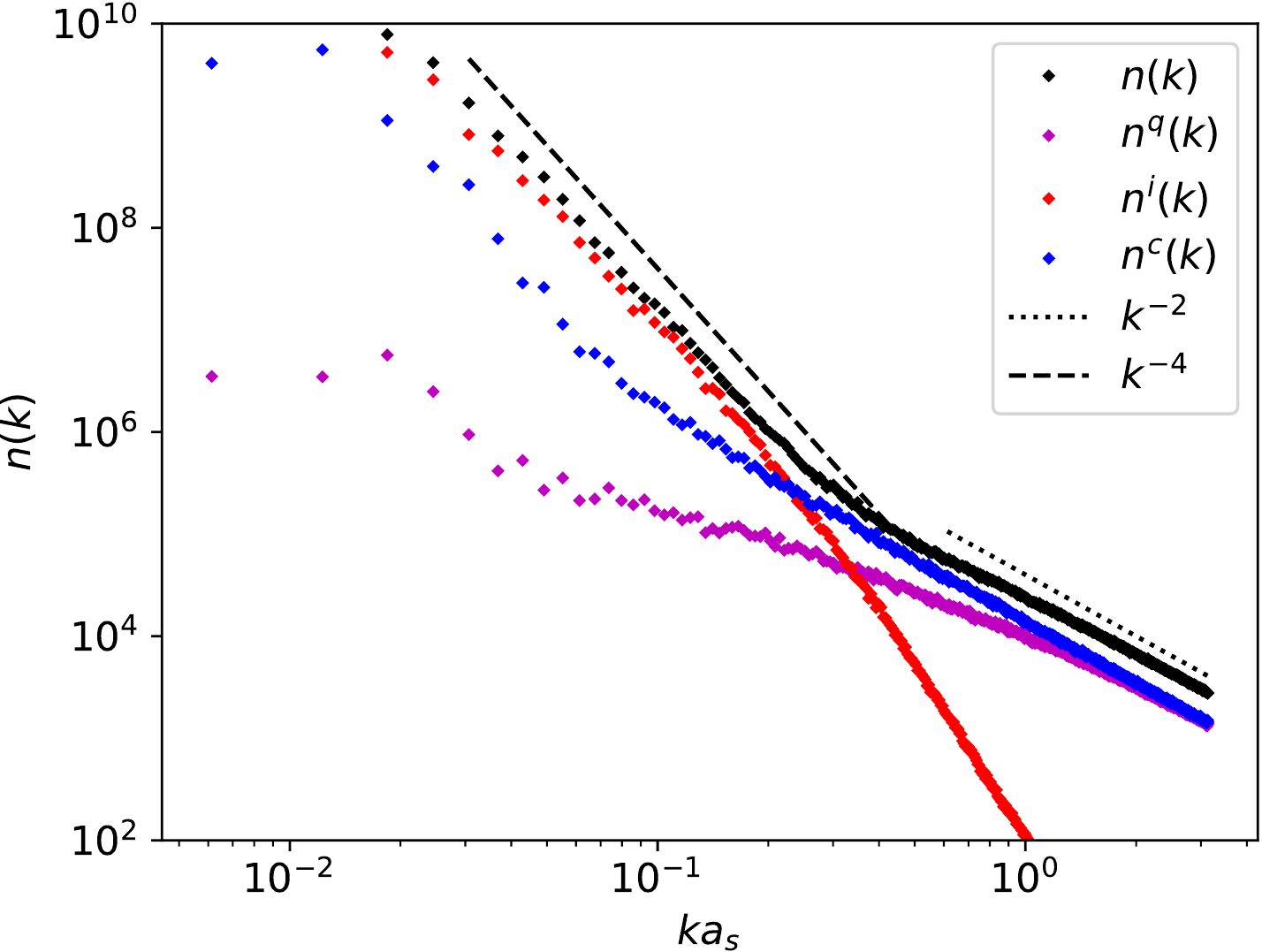}
    \caption{Occupation numbers for different fractions of the system with $\gamma=0.6$ at $t=5\times10^4\xi_d^2$: single particle spectrum for quantum pressure (purple diamonds), incompressible (red diamonds) and compressible (blue diamonds) contributions. 
    The total occupation number (black diamonds) for the single particle spectrum is obtained by summing the corresponding fractions from each condensate component. The single particle spectrum obeys a $k^{-2}$ scaling (dotted line) in the ultra-violet and a $k^{-4}$ scaling (dashed line) in the infrared regions.}
    \label{fig:spectra}
\end{figure}

Fig.~\ref{fig:spectra} shows the occupation number for each energy contribution along with the total occupation number $n(k)$ for the case of $\gamma=0.6$ at a time $t=2 \times 10^5\tau$. The total single-particle spectrum obeys the predicted scaling $n(k) \sim k^{-4}$ in the infrared (IR) region and $n(k) \sim k^{-2}$ in the ultraviolet (UV) seen for some turbulent, 2D, scalar BEC systems~\cite{nowak-PRA-2012,karl-NJP-2017,salman-PRA-2016}.
Decomposing the kinetic energy into its constituent parts, we see that the incompressible contribution dominates in the IR and is responsible for the change in scaling to $k^{-4}$ in this region. This incompressible contribution is associated with the vortices in the system~\cite{karl-PRA-2013}. At large $k$, the spectrum is dominated by the compressible and quantum pressure contributions exhibiting the weak-wave-turbulence scaling $k^{-2}$.  This scaling of the energy is qualitatively insensitive to variations in $\gamma$.

Next, we consider the time-dependent properties of the turbulent dynamics. For this purpose, we will concentrate on the correlation functions for the spin and mass parts of the pseudospinor order parameter. For a homogeneous turbulent system these are defined, respectively, as~\cite{symes-PRA-2017}
\begin{align}
    G_\Phi(\vect{r}, t) &= \frac{2}{n^2}\mathrm{Tr}\left[ \langle \mathsf{Q}(\vect{0})\mathsf{Q}(\vect{r}) \rangle \right] \, , \\
    G_\Theta (\vect{r}, t) &= \frac{1}{n^2} \langle \alpha^*(\vect{0})\alpha(\vect{r}) \rangle \, ,
\end{align}
where $\langle \cdot \rangle$ denotes ensemble averaging.
Here, the matrix
\begin{align}
    \mathsf{Q} = \begin{pmatrix}
       Q_{xx} & Q_{xy} \\
       Q_{xy} & -Q_{xx}
    \end{pmatrix},
\end{align}
where $Q_{xx} = \mathrm{Re}\{ \psi_1^*\psi_2\}$ and $Q_{xy} = \mathrm{Im}\{ \psi_1^*\psi_2\}$, is associated with spin ordering in the system, while 
$\alpha=-2\psi_1\psi_2$ is an alignment parameter.
Exploiting the fact that our turbulent system is homogeneous, we can replace ensemble averages with spatial averages. The spin correlation function is then equivalently defined as~\cite{symes-PRA-2017}
\begin{align}
    G_\Phi(r, t) = \int \upd\Omega_r \int \frac{\upd^2\vect{x}'}{L^2}
    \frac{2\mathrm{Tr}\left[ \mathsf{Q}(\vect{\vect{x}'})\mathsf{Q}(\vect{x}' + \vect{r}) \right]}{n^2},
\end{align}
where $\int \upd\Omega_r$ denotes angular integration. We perform the same averaging for the superfluid correlation function.
\begin{figure}
    \centering
    \includegraphics[width=\columnwidth]{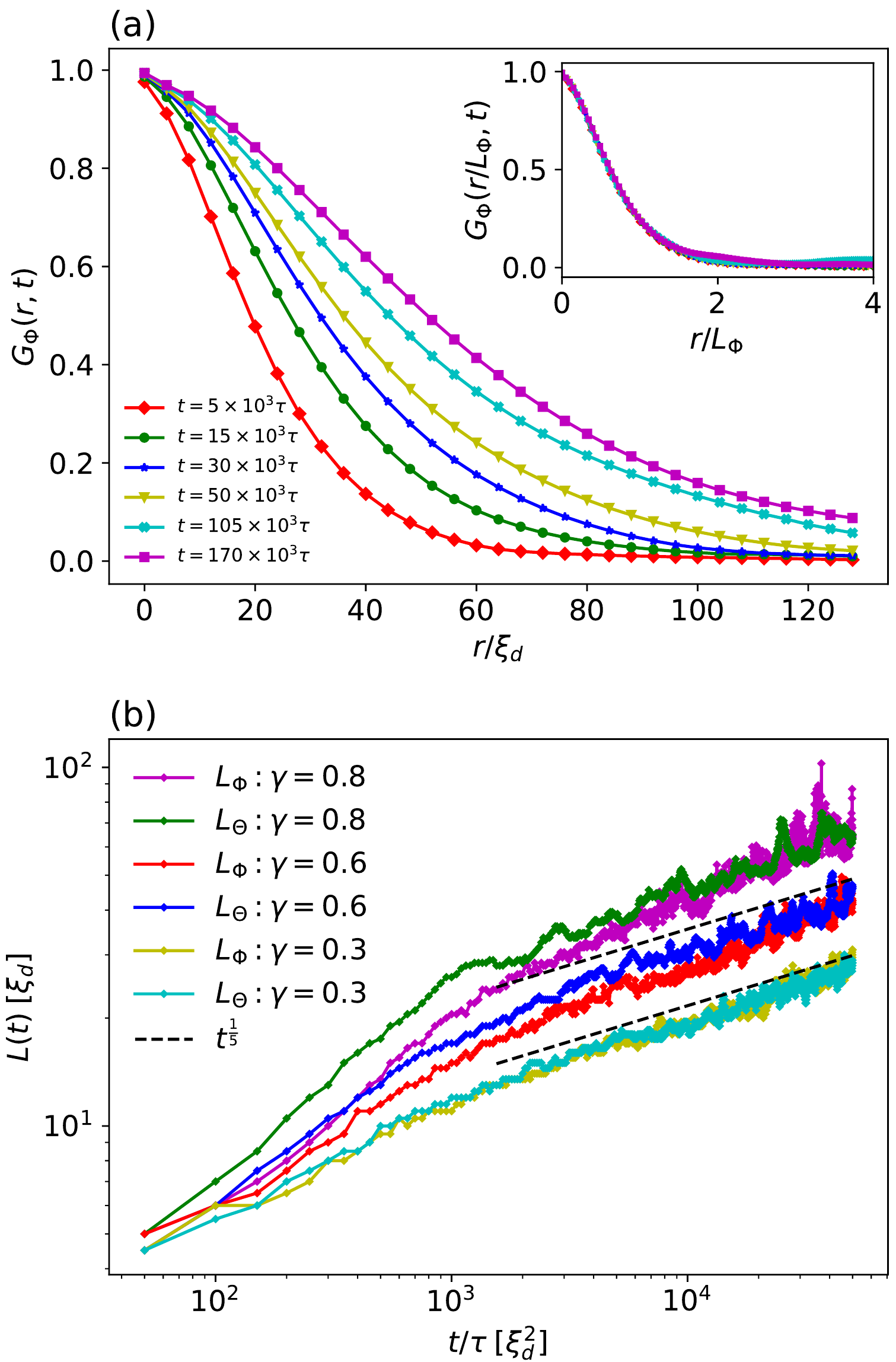}    
    \caption{(a): Spin-correlation function as a function of time for $\gamma=0.6$. The spin order decays more slowly as time increases, indicating domain growth within the system. Inset: collapse of the spin correlation function when scaled by the spin correlation length. (b): Correlation lengths corresponding to the spin and superfluid correlation functions as a function of time for $\gamma=0.3, 0.6, 0.8$. Larger $\gamma$ give a faster initial growth, with a universal scaling appearing for $t \gtrsim 2.5\times 10^3\xi_d^2$. The $t^{1/5}$ scaling predicted from the scalar BEC is indicated for comparison.}
    \label{fig:correlations}
\end{figure}

\begin{figure}[h!]
    \centering
    \includegraphics[width=\columnwidth]{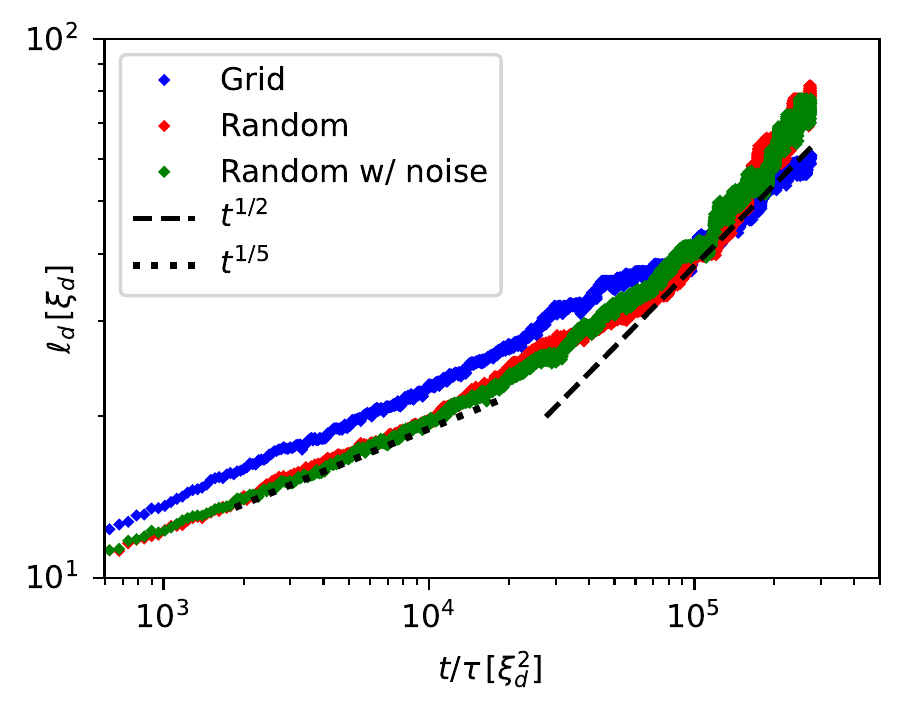}
    \caption{Mean vortex distance in a scalar BEC for three different initial conditions using the same parameters as in ref.~\cite{karl-NJP-2017}. The $t^{1/5}$ early-time as well as the $t^{1/2}$ late-time scaling regimes are recovered.}
    \label{fig:scalar_lengthscale}
\end{figure}

In fig.~\ref{fig:correlations}(a) we plot the results for the spin correlation function at different times for $\gamma=0.6$. As the time increases, the correlation function decays over a larger distance, indicating the emergence of long-range order within the system. We verify the same behaviour for the mass-correlation function. From these correlation functions we obtain the correlation length, $L_\delta(t)$ for $\delta=\{\Phi, \Theta\}$, which we take as the distance at which the corresponding correlation function decays to a quarter of its value at $r=0$: $G_\delta(L_\delta, t) = \frac{1}{4}G_\delta(0, t)$. The correlation functions are said to exhibit dynamical scaling when their form at different times remains self similar. This means that they collapse to a universal, time-independent function when scaled by the correlation lengths, i.e.~$H_\delta(r)=G_\delta(r/L_\delta(t), t)$.
The inset in fig.~\ref{fig:correlations} shows this collapse of the spin correlation function in our system, indicating that $G_{\Phi}(r,t)$ does indeed exhibit dynamical scaling. We again verify the same behaviour for $G_{\Theta}(r,t)$.

Fig.~\ref{fig:correlations}(b) shows both correlation lengths $L_{\Phi,\Theta}(t)$ as a function of time for $\gamma=0.3$, $\gamma=0.6$ and $\gamma=0.8$. After the initial evolution the temporal scaling of the correlation lengths becomes universal for all values of $\gamma$. 
However, the effect of $\gamma$ is apparent in the early time evolution where a larger $\gamma$ leads to a faster growth of the correlation lengths. This is indicative of a difference in the decay rate of the vortices in the early-time dynamics.


\begin{figure*}[h!]
    \centering
    \includegraphics[width=\textwidth]{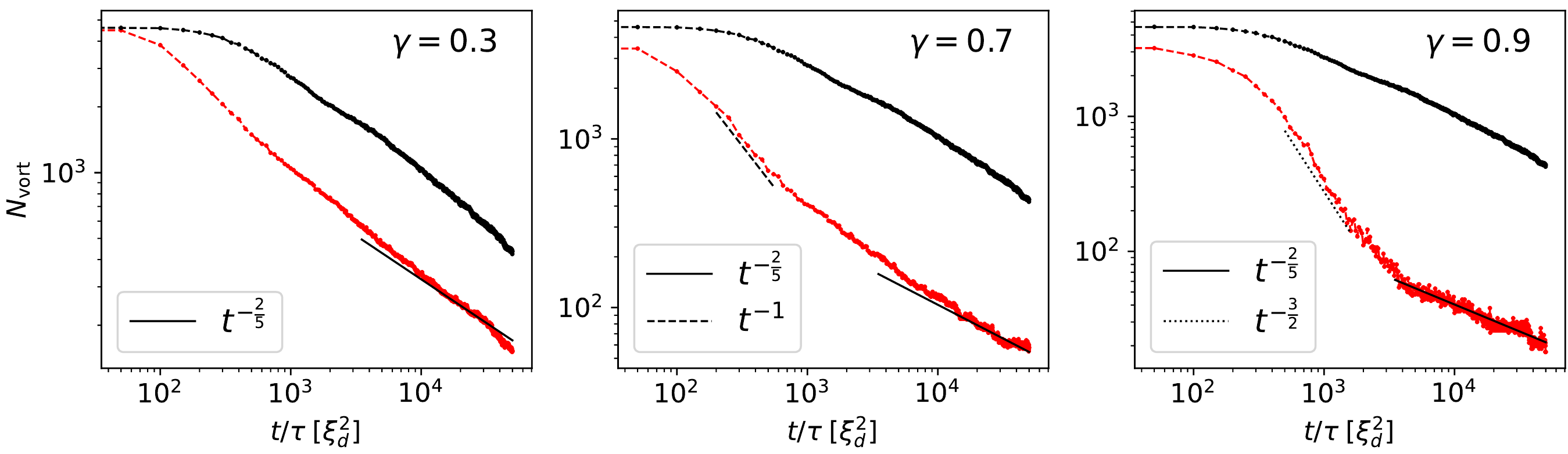}    
    \caption{Total vortex number in both components (red) as a function of time for $\gamma=0.3, 0.7, 0.9$.  Larger $\gamma$ leads to a steeper decay due to the rapid annihilation of opposite-signed vortices in the same component. Overlaid for comparison is twice the vortex number (black) from a corresponding scalar-BEC simulation (equivalent to $\gamma=0$) with the same initial vortex distribution, atom number, and interaction strength $g$ as $\psi_1$.}
    \label{fig:vortex_number}
\end{figure*}

\begin{figure}
    \centering
    \includegraphics[width=\columnwidth]{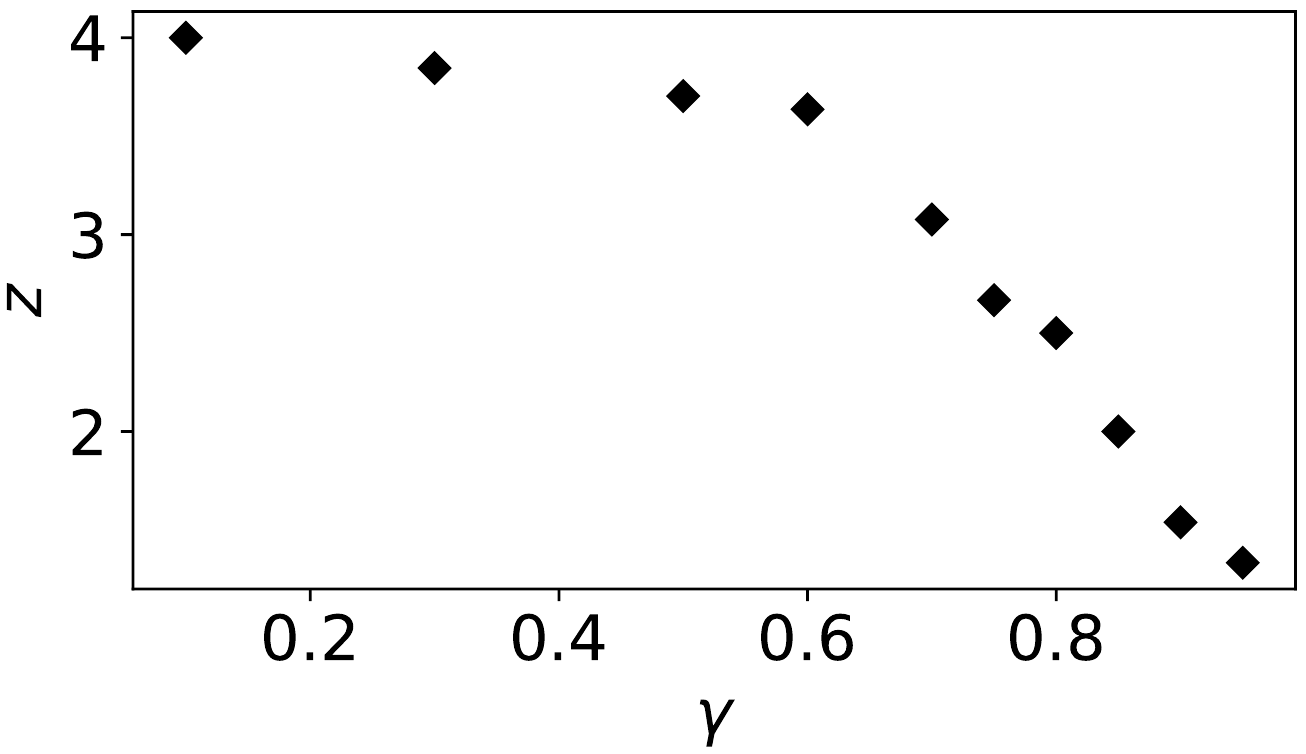}
    \caption{Exponent $z$ as a function of $\gamma$ in the interval $2.5\times 10^2\xi_d^2 < t < 2.5\times 10^3\xi_d^2$. A rapid decrease of the exponent arises for $\gamma\gtrsim0.6$.}
    \label{fig:gamma_vs_expo}
\end{figure}

We can investigate this behaviour by considering the total number of vortices in the system as a function of time. We extract the mean distance between vortices as $\ell_d = 1 / \sqrt{N_\mathrm{vort}}$, where $N_\mathrm{vort}$ is the total number of vortices in the system. As a point of reference, in a scalar BEC initially containing a large number of vortices, $\ell_d \sim t^{\beta}$~\cite{karl-NJP-2017}, where $\beta$ characterises the vortex annihilation rate. In particular, after some (possibly short) period of evolution, a $\beta=1/5$ scaling is observed. For comparison, we have indicated this theoretically expected scaling in fig.~\ref{fig:correlations}(b) for our two-component BEC. At late times, a $\beta=1/2$ scaling appears in the scalar BEC, whose onset is delayed if the initial vortex distribution is highly clustered~\cite{karl-NJP-2017}. In fig.~\ref{fig:scalar_lengthscale} we reproduce this late-time scaling using the parameters of ref.~\cite{karl-NJP-2017} for an initial grid of elementary vortices analogous to our two-component initial state, as well as for a random vortex distribution with and without noise added to the energy spectrum. In all cases we recover both the $t^{1/5}$ scaling after initial evolution and the $t^{1/2}$ late-time scaling, indicating that this behavior is robust and qualitatively insensitive to details of the initial condition.

Motivated by this previous work, we perform a similar analysis to establish how these results extend to a two-component system with HQVs and how the vortex annihilation rate depends on $\gamma$. We focus on the early vortex evolution, where fig.~\ref{fig:correlations}(b) suggests that the $\gamma$-dependence is significant.  Fig.~\ref{fig:vortex_number} shows $N_\mathrm{vort}$ as a function of time for three different values of $\gamma$. For $\gamma=0.7$ and $0.9$, a new scaling regime emerges at early times ($ 2.5\times 10^2\xi_d^2 \lesssim t \lesssim 2.5\times 10^3\xi_d^2$), where $N_\mathrm{vort}(t)$ decays as $t^{-1}$ ($\gamma=0.7$) and $t^{-1.5}$ ($\gamma=0.9$). For $t \gtrsim 2.5\times 10^3\xi_d^2$, $N_\mathrm{vort}(t)$ approaches a universal $t^{-2/5}$ scaling corresponding to $\ell_d \sim t^{1/5}$, similar to the scalar BEC also shown. These results imply a better agreement with the theoretical $t^{1/5}$ scaling than indicated from the correlation lengths [fig.~\ref{fig:correlations}(b)]. 
This suggests that although their growth is driven by vortex annihilation, the length scales $L_{\Phi,\Theta}(t)$  are not fully equivalent to $\ell_d(t)$.
The region of interest in fig.~\ref{fig:vortex_number} only extends up to $t=5\times 10^4 \xi_d^2$ and we therefore expect a universal transition to $\ell_d \sim t^{1/2}$ at times extending beyond the time interval of our simulations.

Previous work has demonstrated that, for a sufficiently high $\gamma \gtrsim 0.6$, a dipole consisting of HQVs with opposite phase winding in the same component will shrink in size as the vortices move toward one another and annihilate~\cite{kasamatsu-PRA-2016}. We therefore attribute the different scaling regime at early times, when the mean inter-vortex separation is small, to this behaviour. This is further supported by the fact that we do not see such scaling for $\gamma\lesssim 0.6$, where such rapid annihilation rate is not prevalent. Within that range of values for $\gamma$, the vortex dynamics begins to recover the behavior observed in a scalar BEC. 

We can model the vortex decay rate by a kinetic-like equation of the form
\begin{align}
    \partial_t N_\mathrm{vort} \sim N_\mathrm{vort}^\eta,
\end{align}
where $\eta>1$.
The dependence of $N_\mathrm{vort}$ on the right-hand side of the equation indicates that the decay rate is a function of the number of vortices that are involved in facilitating the annihilation.
Using this simple model, we can derive temporal scaling of the total vortex number as~\cite{bray-Adv-1994}
\begin{align}
    N_\mathrm{vort} \sim t^{-2/z},
\end{align}
where $z=-2(1-\eta)$. We note that an exponent of $z=2$ corresponds to a two-body collision process whereas $z=5$ corresponds to three-body collisions~\cite{karl-NJP-2017}. In fig.~\ref{fig:gamma_vs_expo}, we quantify the $\gamma$ dependence of the early-time scaling by considering the exponent $z$ in the region $2.5\times 10^2\xi_d^2 < t < 2.5\times 10^3\xi_d^2$. We see a rapid decrease of the exponent after $\gamma>0.6$, when the more rapid annihilation becomes prevalent. The observed decrease in the value of $z$ with $\gamma$ in our simulations signals an additional interaction effect not present in the scalar system.





\section{Conclusions}
We have investigated spatial and temporal aspects of a decaying turbulent two-component BEC containing HQVs. The occupation-number spectrum is found to show a scaling behaviour consistent with similar results for a scalar BEC across a wide range of values of the inter-component interaction strength.

However, we find that a new interaction-dependent scaling regime appears in the temporal properties of the mass- and spin-correlation functions, as well as the mean inter-vortex separation. For large values of the relative inter-component interaction strength, $\gamma \gtrsim 0.6$, these exhibit a $\gamma$-dependent scaling that is markedly different from the universal behavior, which conforms to that of a scalar BEC at a similar stage of time evolution.
Modelling the total vortex number using a simple kinetic equation, we have found that this early-time decay rate for high $\gamma$ cannot be explained by simple two- or three-body collisions. The observed enhanced vortex decay rate at early times for large $\gamma$ may be due to the role played by an additional inter-vortex force that arises between vortices in the same component. The results suggest that this force is short-range and appears in addition to the well-known $1/R$ inter-vortex force. This latter force appears to dominate once the vortex density drops significantly following the rapid vortex annihilations occurring at early times.







\acknowledgments
The results presented were obtained using the High Performance Computing Cluster supported by the Research and Specialist Computing Support service at the University of East Anglia.


\end{document}